\documentclass[useAMS,usenatbib,usegraphicx]{mn2e}

\usepackage{fixltx2e}

\newcommand\Msun  {${\rm M}_\odot$}
\newcommand\msun{\rm M_{\odot}}
\newcommand\mdot  {\dot{M}}

\newcommand\rhoo{\rho_{\circ}}
\newcommand\cmthree{\rm cm^{-3}}

\newcommand\be{\begin{equation}}
\newcommand\en{\end{equation}}
\def\alamenos#1{$^{-#1}$}
\def\ala#1{$^{#1}$}
\def\diezalamenos#1{10$^{-#1}$}

\def\Router{{\rm Router}}
\def\vtot{{v_{\rm tot}}}
\def\tff{t_{\rm ff}}

\def\apj{ApJ}
\def\apjl{ApJL}
\def\aap{A\& A}

\def\araa{ARA\&A}
\def\mnras{MNRAS}

\def\mnras{{MNRAS}}

\title[BHL accretion and the IMF] {Bondi-Hoyle-Littleton accretion and the upper mass stellar IMF} 

\author[Ballesteros-Paredes, et al.]
 { \parbox{7.0in}{Javier Ballesteros-Paredes\ala 1 \thanks{e-mail:
        {\tt j.ballesteros@crya.unam.mx}}, Lee W. Hartmann\ala 2, Nadia P\'erez-Goytia\ala 1,\\
and Aleksandra Kuznetsova\ala 2
} \\
\\
 \ala 1 Centro de Radioastronom\'ia y Astrof\'isica,
            Universidad Nacional Aut\'onoma de M\'exico, \\
            Apdo. Postal 72-3 (Xangari), Morelia,
            Michoc\'an 58089, M\'exico \\
\\
     \ala 2 Department of Astronomy, University of Michigan,  500
           Church Street, Ann Arbor, MI 48105, USA \\
}

\begin{document}

\date{Submitted to MNRAS, \today}

\pagerange{\pageref{firstpage}--\pageref{lastpage}} \pubyear{2012}
 
\maketitle

\label{firstpage}

\begin{abstract}
We report on a series of numerical simulations of gas clouds with
self-gravity forming sink particles, adopting an isothermal equation
of state to isolate the effects of gravity from thermal physics on the
resulting sink mass distributions.  Simulations starting with
supersonic velocity fluctuations develop sink mass functions with a
high-mass power-law tail $dN/d\log M \propto M^{\Gamma}$, $\Gamma = -1
\pm 0.1$, independent of the initial Mach number of the velocity
field.  Similar results but with weaker statistical significance hold
for a simulation starting with initial density fluctuations.  This
mass function power-law dependence agrees with the asymptotic limit
found by Zinnecker assuming Bondi-Hoyle-Littleton (BHL) accretion,
even though the mass accretion rates of individual sinks show
significant departures from the predicted $\mdot \propto M^2$
behavior.  While BHL accretion is not strictly applicable due to the
complexity of the environment, we argue that the final mass functions
are the result of a {\em relative} $M^2$ dependence resulting from
gravitationally-focused accretion.  Our simulations may show the
power-law mass function particularly clearly compared with others
because our adoption of an isothermal equation of state limits the
effects of thermal physics in producing a broad initial fragmentation
spectrum; $\Gamma \rightarrow -1$ is an asymptotic limit found only
when sink masses grow well beyond their initial values.  While we have
purposely eliminated many additional physical processes (radiative
transfer, feedback) which can affect the stellar mass function, our
results emphasize the importance of gravitational focusing for massive
star formation.

\end{abstract}


%%%%%%%%%%%%%%%%%%%%%%%%%%%%%%%%%%%%%%%%%%%%%%%%%%%%%%%%%%%%%%%%%%%%%%%%

\section{Introduction}

The origin of the stellar initial mass function (IMF) has been the
subject of many theoretical studies over the years, with little
resulting consensus.  As discussed in the recent review by
\citet{offner14}, currently popular theories may be classified into
two groups: one in which masses of dense structures or cores created
as a result of (supersonic) turbulence map into the resulting stellar
masses \cite[e.g.,][]{padoan02,hennchab08,hennchab09}, while the other
invokes accretion from an extended region with dynamical interactions
between self-gravitating gas and stars \citep[e.g.,]
[]{bonnell01a,bonnell01b,bate03,clark07}.  Although there seems to be
general agreement that thermal physics plays an essential role in the
turnover of the IMF in the solar-mass (and below) regime
\citep{jappsen05,bonnell06,bate12,krumholz10,myers14}, the application
of Jeans mass-type arguments is far less clear for massive stars,
where continuing accretion from regions beyond the local Jeans lengths
is plausible \citep{bonnell01a}.

One plausible explanation for the upper mass IMF is
gravitationally-focused mass addition, sometimes referred to as
``Bondi-Hoyle'' \citep{zinnecker82,bonnell01a,bonnell01b} or
``Bondi-Hoyle-Littleton'' accretion \citep[BHL;][]{edgar04}, with an
accretion rate given by
\begin{equation}
  \mdot = \frac{4 \pi G^2 M^2 \rhoo}{(c_s^2 + v^2)^{3/2}} \equiv
  \alpha M^2\,,
\label{eq:bhl}
\end{equation}
where $\mdot$ is the mass accretion rate, $M$ is the gravitating
central mass, and $\rhoo$, $c_s$, and $v_{\infty}$ are the ambient gas
density, the (isothermal) sound speed, and the relative velocity of
the ambient gas, respectively.  \citet{zinnecker82} showed
analytically that, if $\alpha$ is constant, a population with a small
but finite difference in initial masses will evolve to an asymptotic
power law slope of $\Gamma = d\log N/d\log M = -1$.  This is a
suggestive result, given the apparent universality (within
observational uncertainties) of the Salpeter slope $\Gamma_s \sim
-1.35$ for high-mass stars \citep{bastian10}, and recognizing that the
$\Gamma \rightarrow -1$ is an asympotic limit reached only when the
accreted masses far exceed initial values \citep{zinnecker82}.

However, the applicability of BHL accretion in massive star formation
has been challenged recently by \citet[][M14]{maschberger14};
\citep[see also][]{padoan14}.  M14 analyzed the results of formation
of sink particles in a hydrodynamic simulation and argued that the
accretion rates are generally $\mdot \propto M^{2/3}$, a result
predicted for the case of accretion in a gas-dominated potential
\citep{bonnell01a,bonnell01b} rather than the the rate of equation
(\ref{eq:bhl}), which applies to accretion in a ``stellar-dominated
potential'' \citep{bonnell01a,bonnell01b}.  M14 found that the sink
accretion follows ``non-linear stochastic processes'', and concluded
that the upper IMF power-law is not a result of BHL accretion.

One of the problems in analyzing the results of numerical simulations
in terms of equation (\ref{eq:bhl}) is that $\alpha$ is not constant,
because the medium surrounding accreting objects is highly structured
and time-variable.  In addition, it is not trivial to relate the
relative velocity in equation (\ref{eq:bhl}), which is presumed to be
that of the ``ambient medium'' far from the (isolated) accreting mass,
with the motions observed in simulations with many gravitating
regions, including both sinks and gas \citep[see, e.g.][]{bonnell01a}.
Nevertheless, if equation (\ref{eq:bhl}) is interpreted as referring
to an average over volume and time, it may still be relevant for the
formation of massive objects, as gravitational acceleration must be
present, especially on large scales.

As numerical simulations of the stellar IMF become increasingly
sophisticated, with ever-more complex physics (e.g., complex equations
of state, radiative transfer, stellar feedback), it becomes
increasingly difficult to isolate the effects of individual
assumptions.  These considerations motivated us to take a simplified
approach. In this paper, we present numerical experiments intended to
address the effects of gravity on the development of upper-mass
power-law IMFs.  We assume isothermality, which allows us to minimize
the importance of (and uncertainties in) thermal physics, and easily
produce sufficient numbers of gravitationally-bound systems (sink
particles) to provide reasonable statistics for the upper-mass IMF
slope $\Gamma$.

Our simulations starting with supersonic but decaying turbulence
produce an upper-mass sink mass function with $\Gamma = -1 \pm 0.1 $;
the run starting with density fluctuations produced a power-law upper
mass tail of comparable slope which emerges at later times.  Thus, we
find the power-law behavior predicted by \citet{zinnecker82}, even
though a simplistic application of equation (\ref{eq:bhl}) with
$\alpha =$~constant, does not reproduce 
the evolving nature of the simulations. We argue however
that these departures from the simple model can be reconciled with the
simulations by taking into account the 
{\it local} 
depletion of matter in the environments of the sinks and the mass
dependence of the time of sink formation.  While our calculations are
of a highly simplified situation compared with that of real
star-forming regions, they nevertheless demonstrate the potential of
gravitational focusing to explain the upper mass power law of the IMF.

\section{Methods}

\subsection{Numerical simulations and initial conditions}

We performed numerical simulations of isothermal gas to represent the
interior of a small ($\sim$ parsec size) molecular cloud.  We used the
$N$-body, smoothed particle hydrodynamics (SPH) code Gadget 2
\citep{springel05} with the inclusion of sink particles as in
\citet{jappsen05}.  Sink particles thus represent stars in the
sense that they do not have gas properties any more, and that the mass
that fell into them will not emerge back into the medium. Since we do
not include radiative feedback or stellar heating, their interaction
with the gas was via gravitational forces only.

Our simulations were performed using 6 million particles, with a total
mass of 1000~\Msun, in a cubic box of side 1~pc in all simulations but
one, which was performed in a box 3~pc on a side.  Sink creation was
allowed above densities of $1.7 \times 10^7 \cmthree$, with an outer
radius of $R_{\rm out} = 1.8 \times 10^{-3}$~pc within which no
further sinks may be made, and an inner sink radius of $R_{in} = 1.8
\times 10^{-4}$~pc or 37 AU.  All runs were isothermal with an adopted
$T = 10$~K.  According to the \citet{bateburkert97} criterion, our
mass resolution limit is $\sim 0.013 \msun$; in practice nearly all of
the sink initial masses were $\ga 0.03 \msun$.  Gas is accreted into
sinks if the particles are within $R_{\rm out}$ and are
gravitationally-bound to the sink, checking previously that another
sink does not preferentially accrete the particle, or if the gas
particle passes within $R_{\rm in}$.  As our sink particles exert
gravity, we impose a gravitational smoothing length of 0.003~pc.

Table~1 shows the main parameters of our runs.  Note that because
these simulations are isothermal, they can be rescaled
\citep[see][]{hsu10}.  We imposed initial velocity fluctuations with a
pure rotational velocity power spectrum with random phases and
amplitudes \citep[e.g.,][]{stone98}, and with a peak at wavenumber
$k_{\rm for} = 4\pi/L_0$, where $L_0$ is the linear size of the box.
No forcing at later times is imposed. All calculations but run 33 were
evolved for $\sim$1.1 free-fall times defined by the initial density
in the computational box (see Table 1).

In the case of run 33, the velocity field was set to zero, and initial
density fluctuations $\delta\rho$ were imposed, following a Kolmogorov
$P_k \propto k^{-11/3}$ spectrum with positive amplitudes only.  We
then stretched these fluctuations such that the density field was set
to
\begin{equation}
  \rho = 10^{(\log{\delta\rho}-{\langle \log{\delta\rho}\rangle})^{1.5}}
\end{equation}

\begin{table*}
 \centering
 \begin{minipage}{140mm}
  \caption{Parameters of the simulations}
   \begin{tabular}{@{}ccccccccc@{}}
  \hline
   Name & Mach \# & Mass (\Msun) & Size (pc) & $n_0 (\rm cm^{-3})$ & $M_J$ ($\msun$) &
   $\tau_{\rm ff} (Myr) $ & $\tau_{\rm end}/\tau_{\rm ff} $ & $\Delta t_{\rm dump}$ (yr)
    \\ \hline 

22   & 8 & 1000 & 1 & 17,000 & 0.48 & 0.24 & 1.17 & 1$\times 10^4$  \\ 
22HR & 8 & 1000 & 1 & 17,000 & 0.48 & 0.24 & 1.08 & 1$\times 10^3$  \\
23   & 16& 1000 & 1 & 17,000 & 0.48 & 0.24 & 1.00 & 1$\times 10^4$  \\ 
24   & 4 & 1000 & 1 & 17,000 & 0.48 & 0.24 & 1.17 & 1$\times 10^4$  \\
32   & 8 & 1000 & 3 & 630    & 2.5  & 1.26 & 1.19 & 5$\times 10^3$  \\ 
33   & 0 & 1000 & 1 & 17,000 & 0.48 & 0.24 & 0.75 & 1$\times 10^4$  \\
\hline
\end{tabular}
\end{minipage}
\end{table*}

Snapshots of the properties of the simulations were taken every
$10^4$~yr, except in the cases of runs 22HR and 32 for which snapshots
were produced every 1000 and 5000~yr, respectively.

We note that all of the runs in Table 1 were actually performed twice,
with the same parameters but with different initial random seeds.  In
the following discussion the figures generally show the results of one
of the two calculations, except for the mass functions where we add
the results of both sub-runs to improve the statistics on the slope of
the mass function.

\subsection{Measuring local properties around the sinks}

Following eq. (\ref{eq:bhl}), we will need to calculate the density
$\rho_0$ and velocity dispersion $v_{\rm turb}$ of the gas around each
sink.  To do so, it is necessary to define some arbitrary radius.  In
principle, it is desirable to calculate these properties outside the
Bondy-Hoyle radius, given by

\begin{eqnarray}
{R_{BH} } & \equiv & {2GM / v_{\rm tot}^2} \nonumber\\ & =& 8.6\times
10^{-3}\ {\rm pc}\ \bigl({M/M_\odot}\bigr) \bigl( {v_{\rm tot} / {\rm
    km\ sec^{-1} } } \bigr)^{-2} ,
\label{eq:bhradius}
\end{eqnarray}
(with $v^2_{\rm tot} = (v^2_{\rm turb} + c_s^2) $ [see eq.
  (\ref{eq:bhl}], $v^2_{\rm turb}$ is the turbulent, or non-thermal
velocity dispersion of the gas around the star, and $c_s$ is the sound
speed).  As can be seen, while the velocity dispersion defines the BH
radius, it in turn depends on the size of the volume at which the
measurement is taken.  Furthermore, since statistically speaking, the
velocity dispersion increases with size, the larger the volume, the
larger the velocity dispersion and the smaller the Bondi-Hoyle radius.
To measure $v_{\rm turb}$, we have, thus, taken a compromise: a volume
large enough to ensure to take measurements at least at 3 times the BH
radius, but small enough to not get a BH radius that is arbitrarily
small.  Thus, we have calculated the velocity dispersion and the mean
density of the sph particles that are within $l=$ 6 times the outer
radius for accretion, {\tt Router}, and verified that $R_{BH} \leq
3\times$~\Router\ in all cases.

\section{Results}

\subsection{Initial velocity perturbations}

\begin{figure}
\centering
\includegraphics[width=0.5\textwidth]{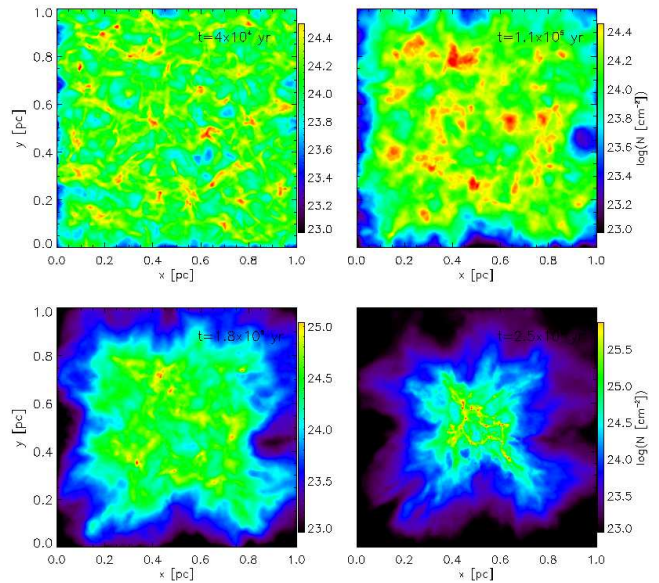}
\caption{Column density of run 22 at times 0.4, 1.1, 1.8 and 2.5
  $\times 10^5$ yr.}
\label{fig:snaps22}
\end{figure}

Figure \ref{fig:snaps22} shows a projection of the evolution of the
density integrated through the computational box, along with sink
particles as they form, at four times for run 22.  The initial
supersonic velocity fluctuations shock and rapidly dissipate energy,
resulting in a (typical) complex distribution of dense clumps and
filaments.  The gas then globally collapses under its own
self-gravity, the density in the clumps and filaments increase,
ultimately forming sink particles which then continue to accrete from
the environment.

\begin{figure}
\centering
\includegraphics[width=0.5\textwidth]{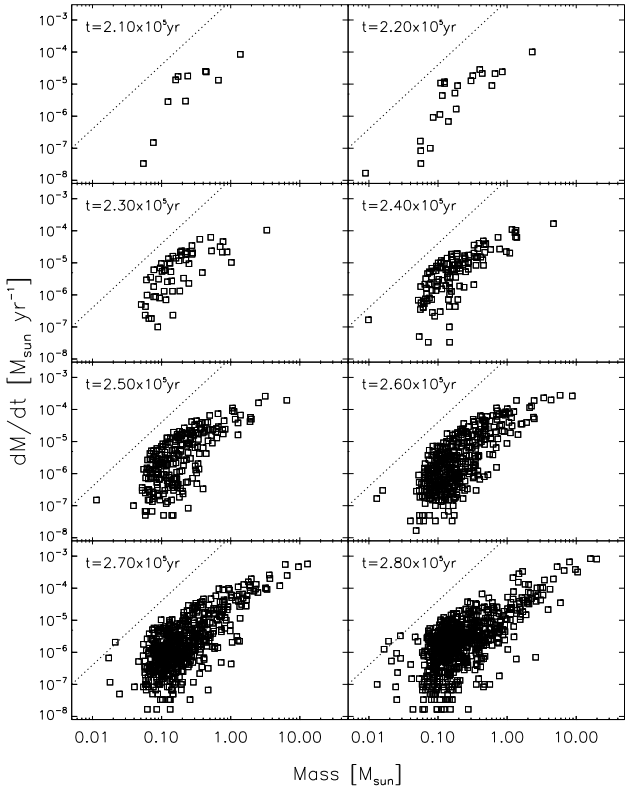}
\caption{Accretion rates into the sinks $\mdot$ {\it vs.} the mass of
  the sinks $M(sink)$ for run 22 from $t=2.1$ to $2.8 \times 10^5$ yr.
  The dotted line indicates a slope $dM/dt \propto M({\rm sink})^2$.}
\label{fig:mdot22}
\end{figure}

We calculated accretion rates into sinks $\mdot$ as the difference in
sink mass between two snapshots, as a function of mass. In Figure
\ref{fig:mdot22} we show the time sequence of accretion rates of sinks
as a function of mass ($M({\rm sink})$).  While the dependence of the
accretion rates on sink mass is crudely comparable to the $\mdot
\propto M^2$ prediction of the simple model, there are departures.  At
the low-mass end of the mass function ($M({\rm sink}) \sim 0.1
\msun$), there is a large spread in the accretion rate at a given
mass, while at the high mass end, the spread in accretion rate {\it
  vs.}\ mass is much less, but the slope is flatter than $M^{-2}$,
with an approximate $M^{1.5}$ dependence at the end of the simulation
for the high-mass sinks.  These departures from the prediction of
equation (\ref{eq:bhl}) are due to the reduction of matter in the
environments of the sinks as mass is accreted, and the time dependence
of sink creation as a function of mass (see discussion in \S 4).  The
low-mass sinks are particularly affected by ``competitive accretion''
\citep[e.g.,][]{bonnell01a,bonnell01b}, where mass is gravitationally
attracted to other sinks, particularly the higher-mass ones.  This
competition results in more low-mass sinks exhibiting very low
accretion rates at later times.

\begin{figure}
\centering
\includegraphics[width=0.5\textwidth]{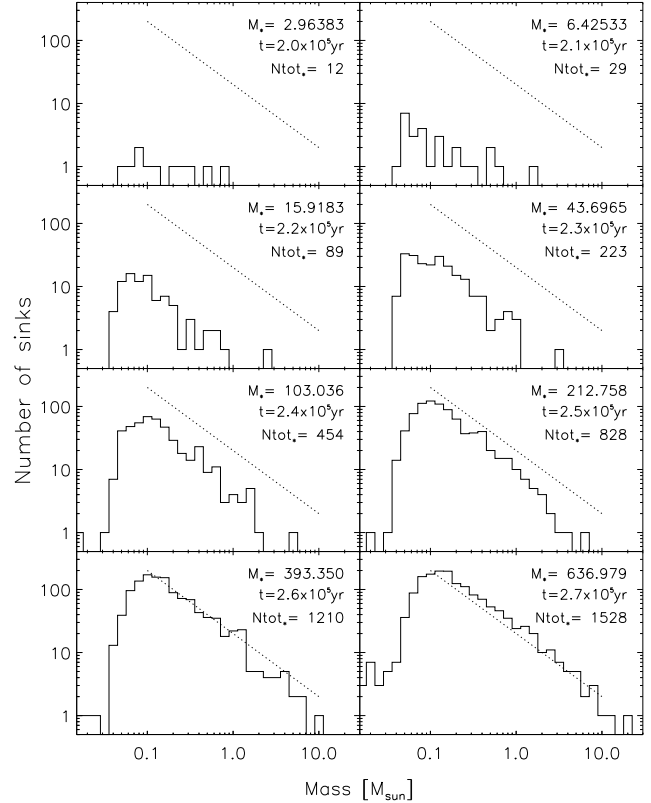}
\caption{The IMFs for run 22, summed over the two random
  initializations, from $t=2.0$ to 2.7$\times 10^5$ yr (compare with
  Figure \ref{fig:mdot22}).  The dotted line indicates a power law
  slope $\Gamma = -1$.}
\label{fig:imfs22}
\end{figure}

Figure \ref{fig:imfs22} shows the evolution of the sink mass function
as a function of time for run 22 (a summation of two individual runs
with different random seed initializations for the phases and
amplitudes of the velocity fluctuations).  Interestingly, a
quasi-power law distribution of masses is established quite early.
Most of the sinks are formed at low masses $\sim 0.05 - 0.1 \msun$.
The slope of the power-law is roughly $\Gamma \sim -1$ over most of
the period of sink formation, but it is not until the end of the
calculation that the statistics become good enough to calculate it
accurately.  Measuring from a low mass of $0.2 \msun$ to the upper
limit yields $\Gamma = -1.12 \pm 0.07$ or from $0.5 \msun$, $\Gamma =
-1.15 \pm 0.10$.

\begin{figure}
\centering
\includegraphics[width=0.5\textwidth]{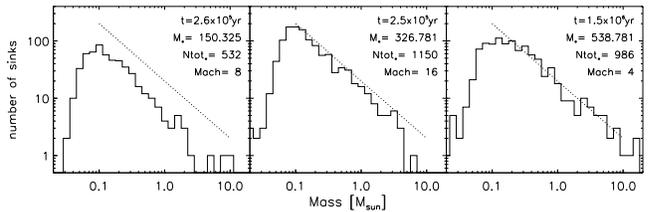}
\caption{The final IMFs for individual runs 23, 24, and 32.} 
\label{fig:imfsall}
\end{figure}

In Figure \ref{fig:imfsall} we show the final IMFs for the four runs
23, 24, and 32.  are $\Gamma = -1.24 \pm 0.30$, $-1.04 \pm 0.07$, and
$-0.96 \pm 0.10$ for runs 23, 24, and 32, respectively, measuring from
the $0.5 \msun$ bin, with essentially the same results measuring from
$0.2 \msun$.  The first three runs demonstrate that changing the Mach
number of the initial velocity fluctuations makes essentially no
difference to the final results.  Run 32 consists of the same mass
within a volume 27 times larger and thus an initial density that
factor smaller; the resulting IMF is identical within errors to the
others.  This demonstrates the scale-free nature of the problem, with
the peak in the mass function controlled by the basic resolution of
these isothermal simulations.  Note that we achieve a $\Gamma \sim -1$
power-law mass function despite the departures of the accretion rates
from a strict $\mdot \propto M(sink)^2$ relation (\S
\ref{sec:BHLrelevant}).

\begin{figure}
\centering
\includegraphics[width=0.5\textwidth]{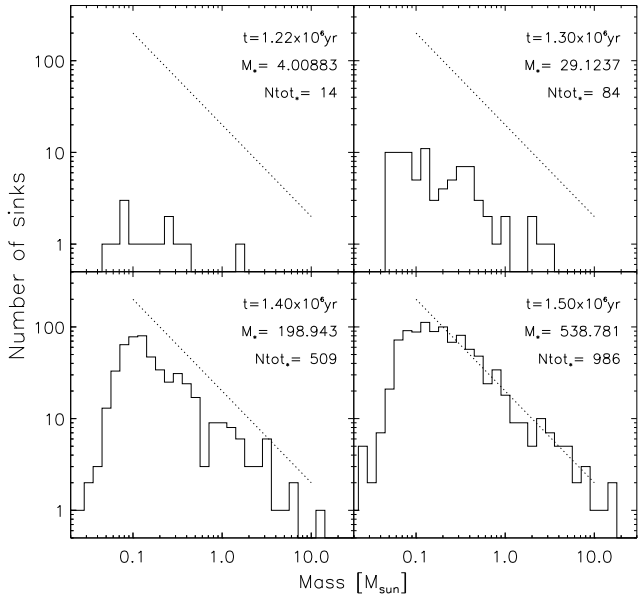}
\caption{IMFs for run 32 at $t=1.22$, 1.3, 1.4 and 1.5 $\times 10^6$.}
\label{fig:imfs32}
\end{figure}

In Figure \ref{fig:imfs32} we use run 32 to explore the onset of the
sink mass function in more detail.  As the initial free-fall time of
this simulation is a factor of 5.2 longer than in the other runs, and
snapshots were dumped every 5000 yr, this simulation effectively
provides higher time resolution (recall that the isothermal nature of
the simulations permits straightforward rescaling).  With higher
(effective) time resolution, the early development of a quasi-power
law distribution of sink masses is observed clearly.

\begin{figure}
\centering
\includegraphics[width=0.5\textwidth]{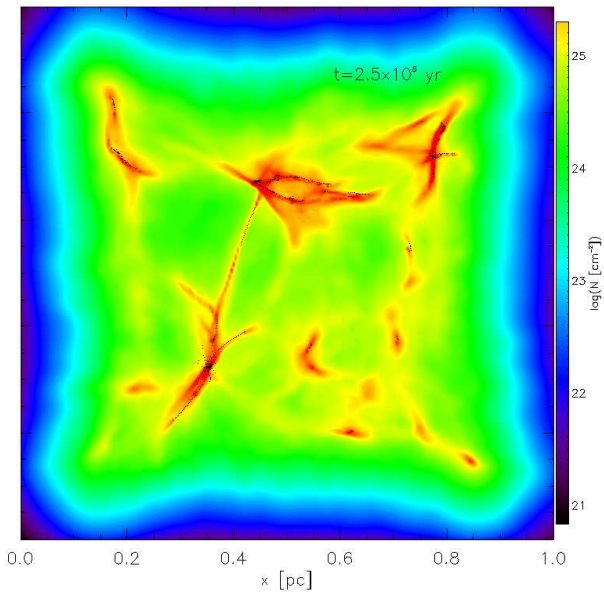}
\caption{Column density at $t=2.5\times 10^5$yr for run 33. Black dots
  denote the current positions of the sinks.}
\label{fig:snaps33}
\end{figure}

\subsection{Initial density fluctuations}

Figure \ref{fig:snaps33} shows the surface density for run 33 at
$t=$0.4, 1.1, 1.8 and 2.5 $\times 10^5$yr.  The adopted initial
conditions are such that the density fluctuations are mostly confined
to a few large-scale concentrations at early times.  The ensuing
gravitational collapse results in the formation of much larger-scale
filaments, with much more linear collections of sink particles than in
the velocity fluctuation cases (Figure \ref{fig:snaps22}).

\begin{figure}
\centering
\includegraphics[width=0.5\textwidth]{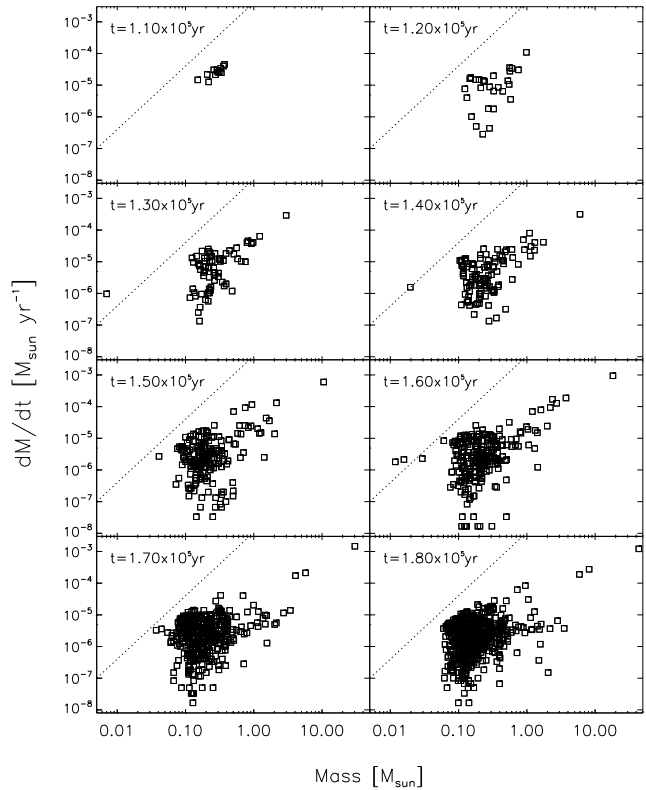}
\caption{$\mdot$ {\it vs.} $M(sink)$ for run 33 from $t=1.1$ to 1.8
  $\times 10^5$ yr.}
\label{fig:mdot33}
\end{figure}

In Figure \ref{fig:mdot33} we show $\mdot\ vs.\ M(sink)$ at various
times.  The accretion rates show considerably less correlation with
sink mass than for the velocity fluctuation case (Figure
\ref{fig:mdot22}).  This difference is probably due to the differing
spatial distributions of the sinks, especially at early times.  In the
density fluctuation case, the dense regions initially are far more
spatially concentrated into narrow filaments, resulting in higher
concentration of sinks and thus more competition for accretable mass,
causing more scatter in accretion rates at a given mass (compare
Figure \ref{fig:snaps33} to Figure \ref{fig:snaps22}).
 
\begin{figure}
\centering
\includegraphics[width=0.5\textwidth]{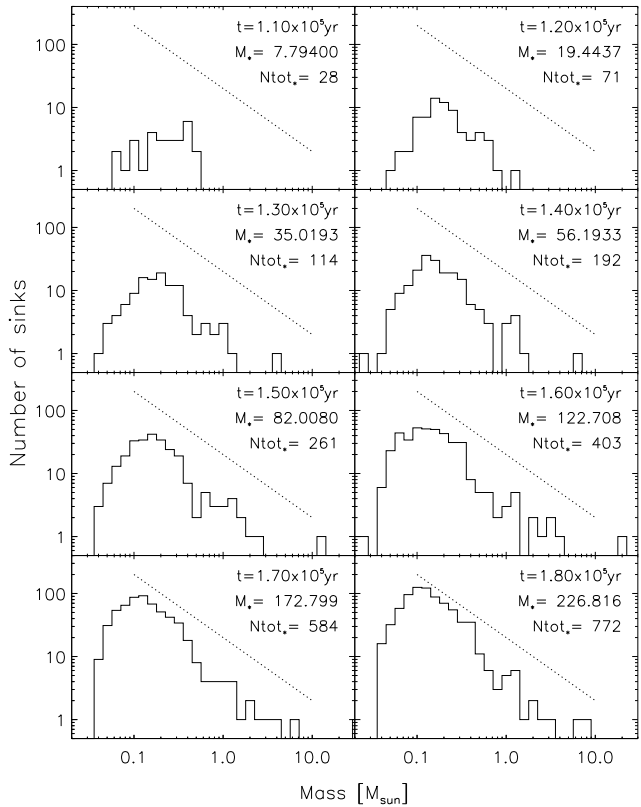}
\caption{IMFs for density-fluctuation run 33 from $t=1.1$ to 1.8
  $\times 10^5$ yr.}
\label{fig:imfs33}
\end{figure}

Figure \ref{fig:imfs33} shows the development of the sink mass
function for run 33. The evolution differs from that of the velocity
fluctuation cases in that a quasi-log-normal distribution emerges
initially, consistent with the weaker correlation of accretion rates
with mass.  However, even in this case, a roughly power-law tail
develops at later times, with an evolution arguably toward $\Gamma
\sim -1$, although the statistics are not good enough to make an
accurate slope measurement.

\section{Discussion}

\subsection{Is BHL accretion relevant?\label{sec:BHLrelevant} }

The standard BHL accretion equation (\ref{eq:bhl}) does not apply
directly to our simulations because it envisages a well-defined
external medium from which to accrete which is unperturbed by any
gravitational field other than that of the central mass.  This is
manifestly not the case because the self-gravity of the gas is
important and even dominant early on, and many other gravitating
objects are present.  Nevertheless, the development of the power-law
mass function with $\Gamma = -1$ as predicted suggests that some basic
feature of BHL accretion dominates the production of the mass function
even in the more complex situation.

In the simple BHL model, using equation (\ref{eq:bhl}) and assuming
$\alpha =$~constant, the growth of the accreting mass as a function of
time is

\begin{equation}
M = \frac{M_0}{1 - \alpha M_0 t}\,,
\label{eq:masstime}
\end{equation}
where $M_0$ is the mass at $t = 0$.  From this, \citet{zinnecker82}
showed that the resulting mass function $\zeta = d N/d \log M$ for an
initial population of masses $M_{0,i}$ with a mass distribution
$\zeta_0 = \zeta(t=0)$ is of the form

\begin{equation}
\zeta(M) ~=~ \zeta_0 \left ( \frac{M}{1 + \alpha M t} \right )
( 1 ~+~ \alpha M t)^{-1}\,.
\label{eq:zinneckerzeta}
\end{equation}

More generally, recognizing the environment of the sinks is both
temporally and spatially variable, one may write the accretion rate as
\begin{equation}
\mdot = \alpha(t,{\bf r}) M^n\,.
\label{eq:simple}
\end{equation}
Even if $\alpha$ is not constant, if it does not depend on $M$ either
explicitly {\em or implicitly}, the analysis of the Boltzmann
continuity equation made by \cite{zinnecker82} is applicable, and the
asymptotic mass function $\zeta = d N/d \log M$ has the form

\begin{equation}
\zeta(M) \propto {d \log M_0 \over d \log M_f} \propto
M^{-(n-1)}\,, (n > 1)\,.
\label{eq:gamma}
\end{equation}
While the mass function slope $\Gamma = -1.0 \pm 0.1$ that we observe
in our velocity-fluctuation simulations indicates $n=2$, the snapshots
themselves naively suggest accretion rates for the most massive sinks
with a shallower mass dependence $n \sim 1.5$ (e.g., Figure
\ref{fig:mdot22}) which would imply $\Gamma \sim -0.5$.

\begin{figure}
\centering
\includegraphics[width=0.5\textwidth]{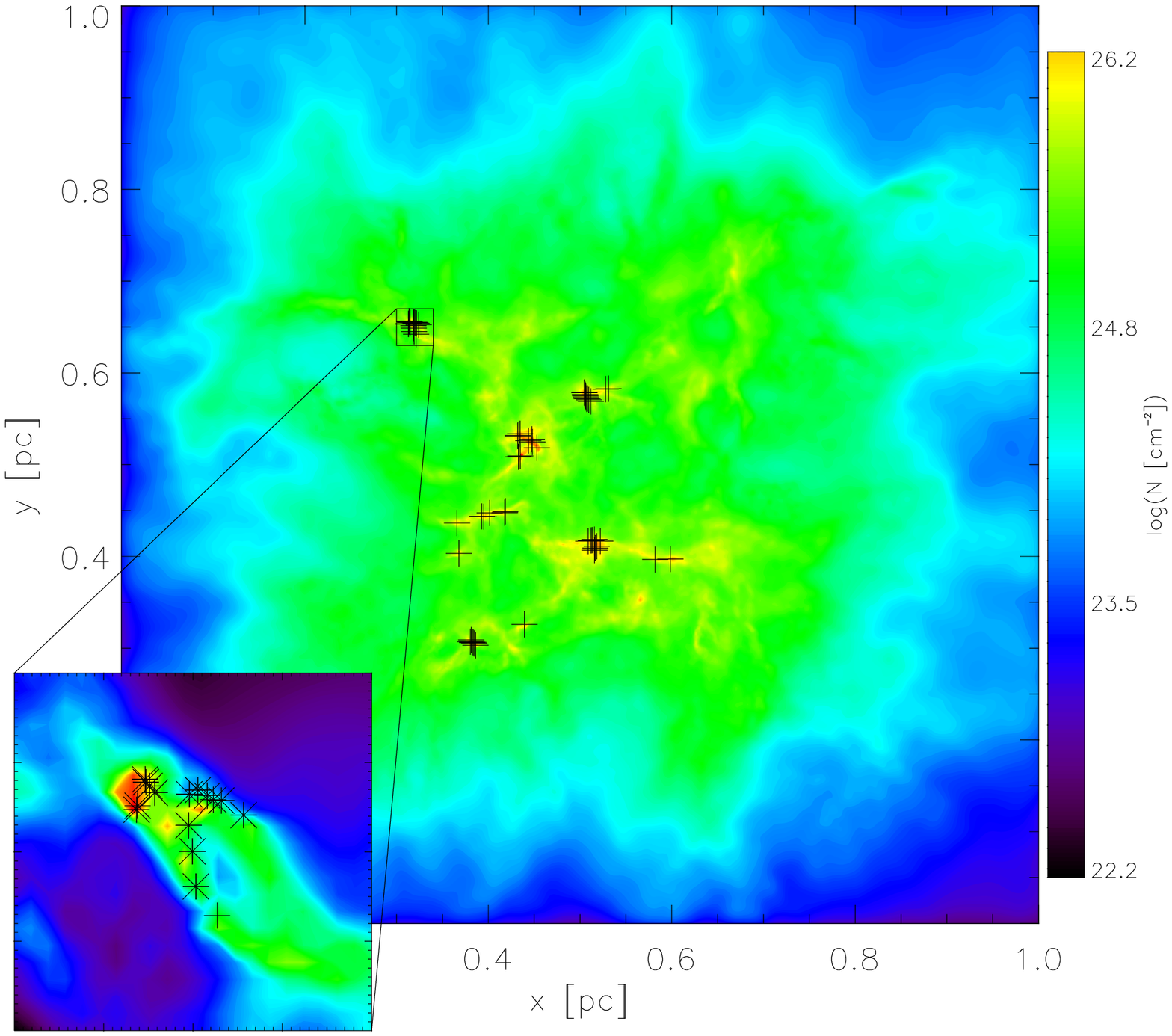}
\includegraphics[width=0.5\textwidth]{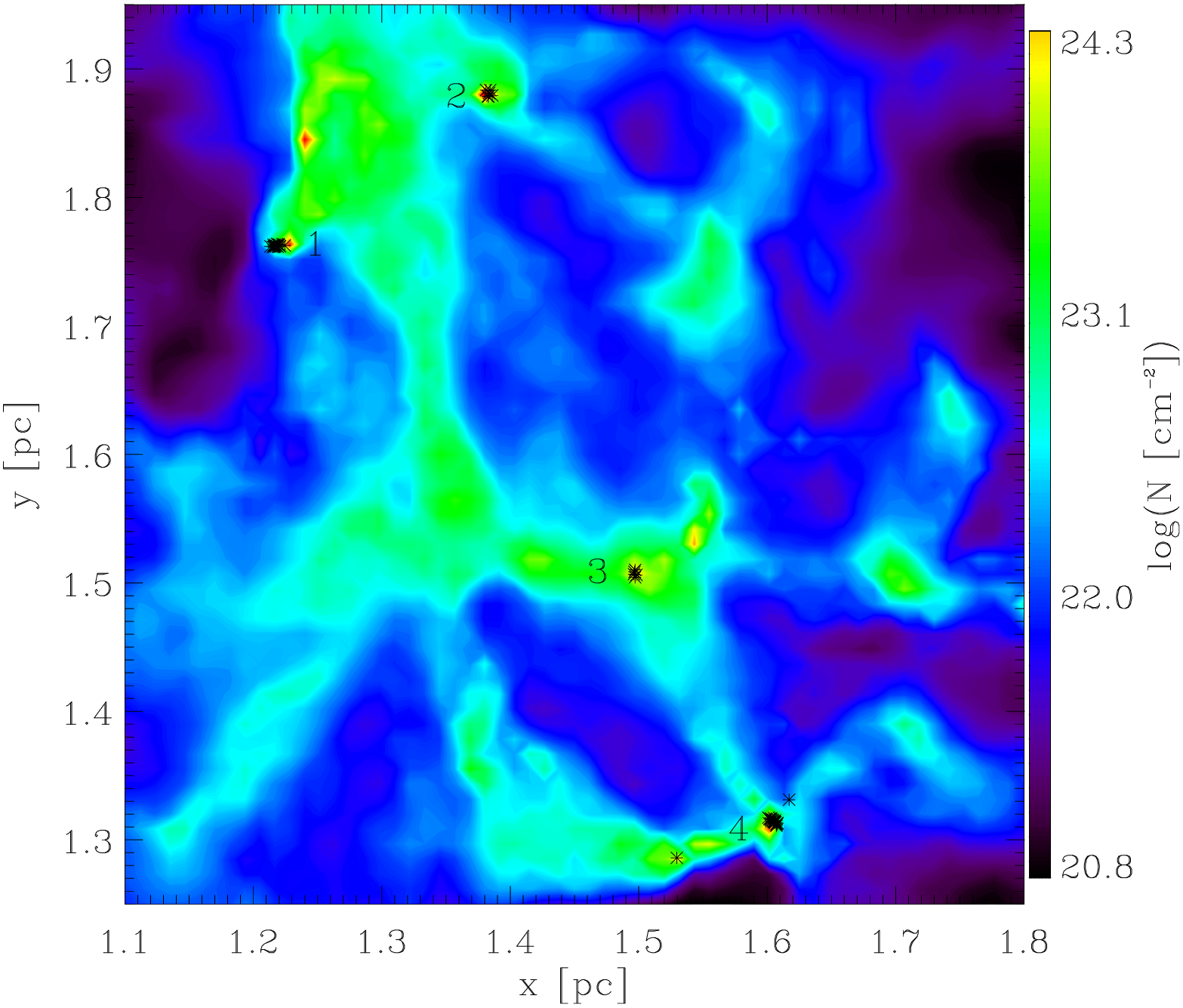}
\caption{Maps of runs {\tt 22ID2} ({\it a}, upper panel) and {\tt 32}
  ({\it (b)}, lower panel) at $t= 220,000$ yr and at $1.325$~Myr,
  respectively.  In both cases sinks are formed by groups, where
  apparently, the local density and velocity dispersion of the gas is
  nearly the same for all the sinks in the group. Closer examination
  shows that this is not necessarily the case (see text).}
\label{fig:snapzoom}
\end{figure}

It is important to recognize that the mass accretion rates of the
sinks, integrated over time, {\em must} map into the mass function.
To resolve this apparent contradiction, we notice, first of all, that
sinks are typically formed in groups: in Fig.  \ref{fig:snapzoom} we
show a snapshot of {\tt run 22ID2} (upper panel) and {\tt 32} (lower
panel) at relatively early times, i.e., when only 64 and 50 sinks have
been formed, respectively ($t= 220,000$ yr, equivalent to
0.91~free-fall times ($t_{\rm ff}$) for {\tt run 22ID2}, and
$1.325$~Myr, equivalent to 1.05 $t_{\rm ff}$ for {\tt run 32}).  Thus,
at first glance it could be arguable that such groups share similar
environmental properties (density and velocity dispersion).

\begin{figure}
\centering
\includegraphics[width=0.5\textwidth]{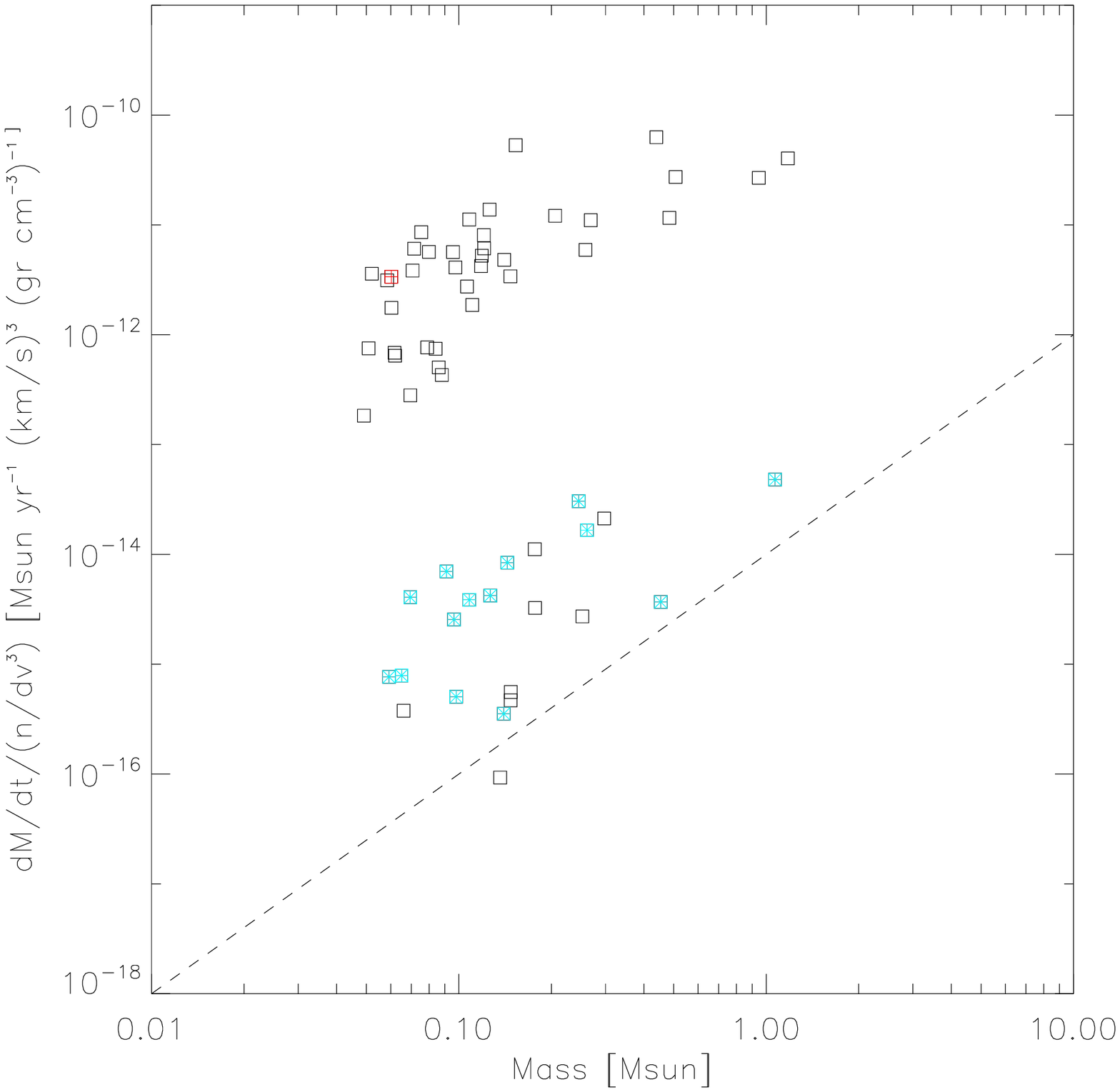}
\includegraphics[width=0.5\textwidth]{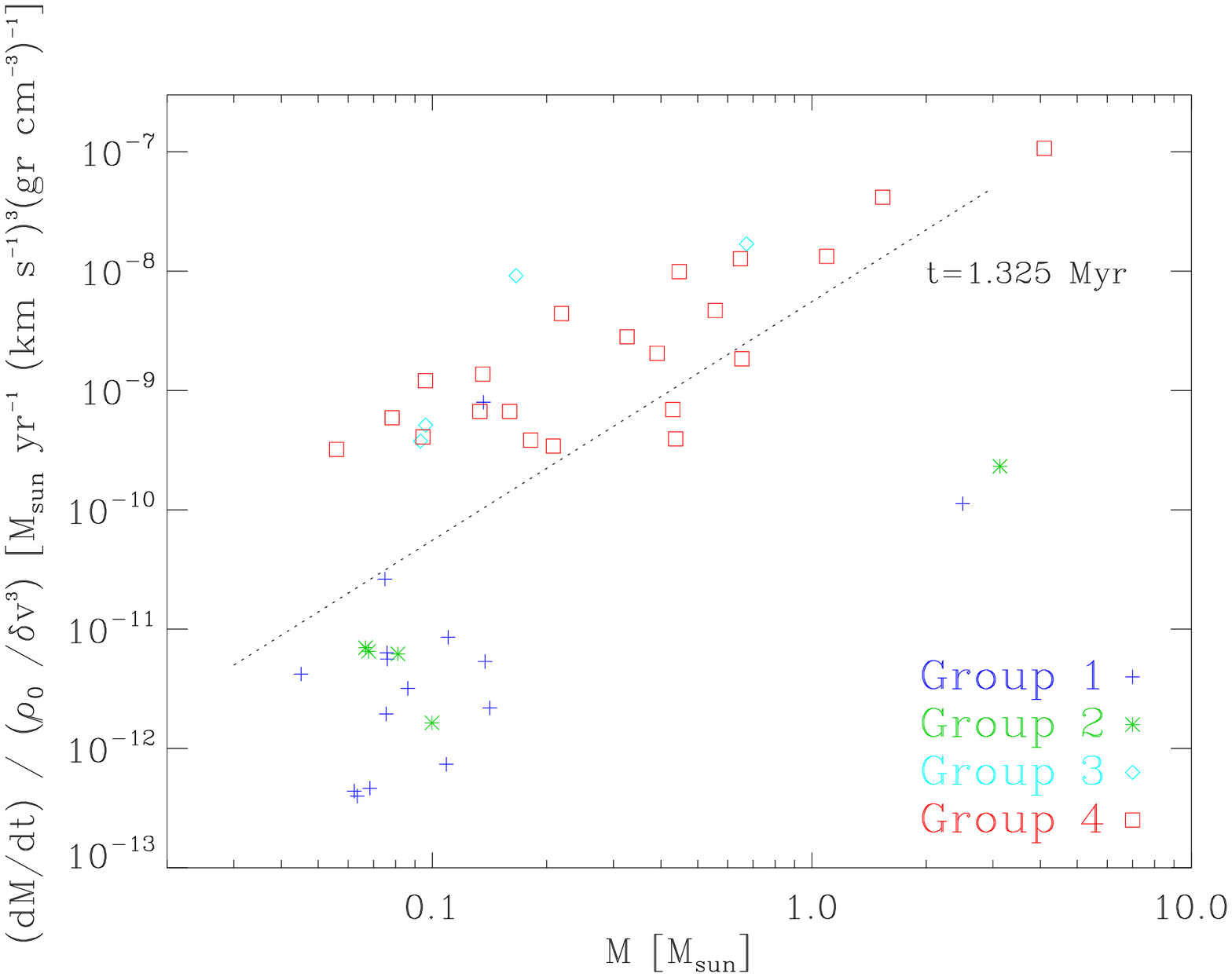}
\caption{$\dot{M}/(\rho v^3)$ vs. $M$ for {\it a}) {\tt run 22ID2},
  upper panel) and {\it b} {\tt run 32} at the same times shown in
  Fig.~\ref{fig:snapzoom}.  The dashed line has a slope of 2.  Note
  that when one accounts for the local variations of $\alpha^\prime$,
  even though there is some scatter, individual groups exhibit nearly
  the quadratic dependence of $\dot{M}$ with $M$.  The groups labeled
  in the lower panel are the same than those shown in the lower panel
  of Fig.~\ref{fig:snapzoom}.}
\label{fig:dmdt_vs_M_Malpha_early}
\end{figure}

We also note that, since $\alpha = \alpha(t, {\bf r})$, local
variations in $\alpha$ may blur any quadratic dependence of $\dot{M}$
on $M$ at a given time.   Thus, by dividing the mass accretion
  rate of each sink by its local $\alpha$, it might be possible to
  recover any power-law dependence (eq. [\ref{eq:simple}]) of the
  accretion mass rate with mass, at least by groups with similar
  environmental properties (i.e., similar $\alpha$):

\begin{equation}
  \log{\dot{M}} = \log{\alpha_i(t, {\bf r})} + n \log{M},
\label{eq:logMdot_eq_logalpha+2logM}
\end{equation}
In Fig.~\ref{fig:dmdt_vs_M_Malpha_early} we now plot the
ratio\footnote{In what follows, since $\alpha\propto \rho/\vtot^3$ we
  will define $\alpha^\prime \equiv \rho/\vtot^3$}
$\dot{M}/\alpha^\prime$ against the mass $M$ for all the sinks at the
same times shown in Fig.~\ref{fig:snapzoom}.  In both cases we notice
that there are two main groups, each one sharing the approximate
tendency $\dot{M} \propto M^2$, but with an offset of $\sim$3 orders
of magnitude between them.  This result shows that different groups of
sinks might have different environmental properties, and thus, any
possible relationship of the kind $\dot{M} \propto M^2$ might be
blured when the groups are analyzed all together.

To worsen the case, we furthermore notice that even when one can
account for such kind of segregation by $\alpha_i$, the situation
might still not be straightforward.  For instance, in the upper panel
of Fig.~\ref{fig:dmdt_vs_M_Malpha_early} (which corresponds to {\tt
  run 22ID2}) we denote by empty squares all the sinks at $t=1.03\tff$
but those that belong to the group that it is zoomed in in the upper
panel of Fig.~\ref{fig:snapzoom}.  The 16 sinks that belong to that
group are denoted also with cyan asterisks if the value of
$\alpha^\prime$ is smaller than \diezalamenos{13}, and by red crosses
otherwise.  We then notice in the upper panel of
Fig.~\ref{fig:dmdt_vs_M_Malpha_early}, that all but one sinks that
belong to the mentioned group have $\alpha^\prime < 10^{13}$, and that
only one of them (red symbol) has $\dot{M} / \alpha^\prime > $
\diezalamenos{13}. This sink is also denoted by a cross in the zoom of
the upper panel of Fig.~\ref{fig:snapzoom} (the remaining ones are
denoted by asterisks), and even though it is the neighborhood of the
group, in the upper panel of Fig.~\ref{fig:dmdt_vs_M_Malpha_early} it
falls far away from the $\dot{M}/\alpha^\prime \propto M^2$ tendency
delineated by the rest of the group.  The reason might be either that
the local density of this sink is smaller than that of the rest of the
group (e.g., it is located in the periphery of the core), or that
somehow this sink has a larger velocity (e.g., the sink its been
ejected), compared to the rest of the group, increasing the velocity
dispersion of the sph particles of its neighborhood.  In fact, a
careful analysis of the zoom in the upper panel of
Fig.~\ref{fig:snapzoom} shows that it might be the first case: the
sink is located in the periphery of the core, according to the column
density map.

In a similar way, we notice that each one of the four groups of sinks
shown in the lower panel of Fig.~\ref{fig:snapzoom}, also exhibit the
approximate $\dot{M}\propto M^2$ tendency, although with some scatter,
(see lower panel of Fig.  ~\ref{fig:dmdt_vs_M_Malpha_early}), and even
with some outliers (e.g., blue cross in group 1 showing
$\dot{M}/\alpha^\prime \sim 5-8\times$\diezalamenos{10}).  These
results suggest that even within a single group, there migh be
different values of $\vtot$ and $\rho$, and then, strong scatter in
the plane ($M$, $\dot{M}/\alpha^\prime$) migh be seen.

\begin{figure}
\centering
\includegraphics[width=0.5\textwidth]{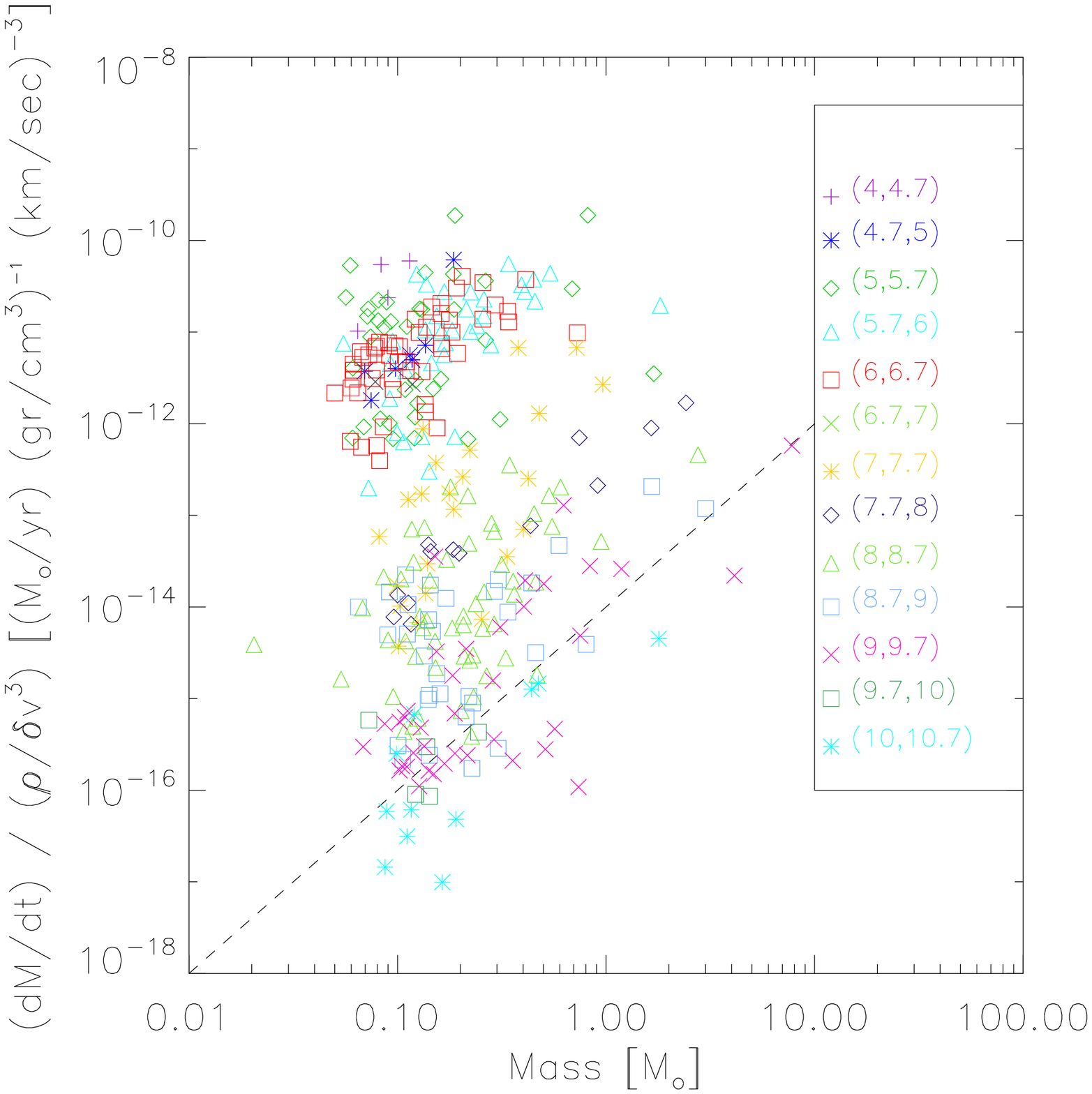}
\includegraphics[width=0.5\textwidth]{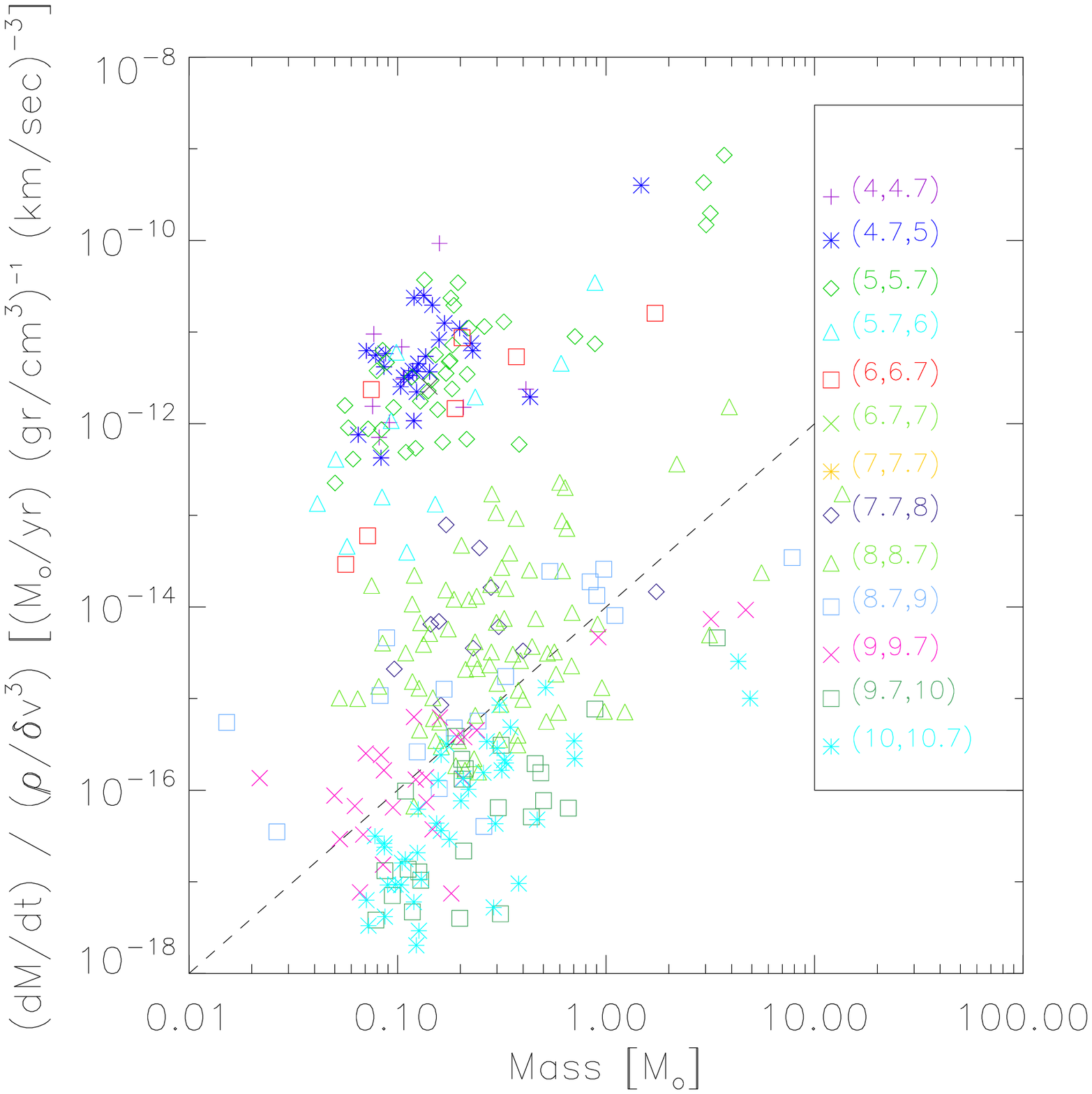}
\caption{$\dot{M}/\alpha^\prime$ vs. M for runs {\it a} {\tt 22ID2} at
  $t= 1.03 \tff$ and {\it b} {\tt 32} $t=1.17 \tff$.  The dashed line
  has a slope of 2, and the actual values of $\alpha^\prime$ are
  labeled and colored accordingly in the rectangle, in units of [cm km
    sec\alamenos 1]\alamenos 3.  Note that when one accounts for the
  local variations of $\alpha^\prime$, individual groups exhibit the
  quadratic dependence of $\dot{M}$ with $M$.}
\label{fig:dmdt_vs_M_Malpha_late}
\end{figure}

To further stress this point, in Fig.~\ref{fig:dmdt_vs_M_Malpha_late}
we plot again $\dot{M}/ \alpha^\prime$ {\it vs.} $M$ for the same two
runs, but (a) at later times, when substantially more sinks have been
formed (363 for {\tt run 22ID2}, upper panel, and 354 for {\tt run
  32}, lower panel), and then the density and velocity fields are even
more complex, and (b) in both cases, we color and label every point
according to the local value of $\alpha^\prime$ (the actual values are
colored and labeled in log scale in the inner box).  Although there is
some scatter at every $\alpha^\prime$, it is clear that in each case
we recover the quadratic dependency per groups, strongly suggesting
that, indeed, the accretion is of Bondi-Hoyle type, with non-constant
$\alpha$, as would be predicted by
eq. (\ref{eq:logMdot_eq_logalpha+2logM}) for $n=2$ and different
$\alpha$.  In summary, {\em even for a single group, substantial
  scatter in the plane ($M$, $\dot{M}/\alpha^\prime$) can be due to
  local fluctuations in velocity or density. These fluctuations blur
  the quadratic dependence of $\dot{M}/\alpha^\prime$ with $ M$.}

\begin{figure}
\centering
\includegraphics[width=0.45\textwidth]{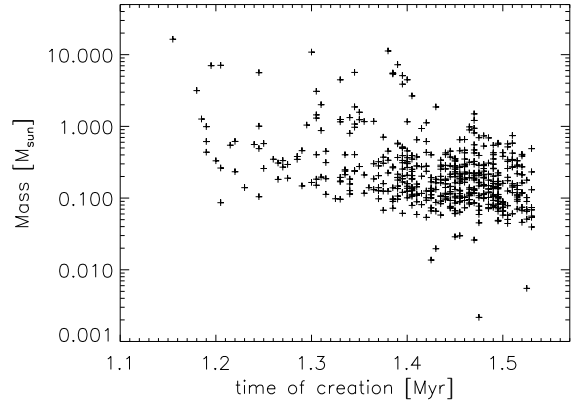}
\caption{Time at which sinks initially form for Run 32.  The sinks
  with larger final masses systematically accrete for longer than the
  lowest-mass sinks (see text).  Similar plots are found for all the
  runs.}
\label{fig:sinktime}
\end{figure}

\begin{figure}
\centering
\includegraphics[width=0.5\textwidth]{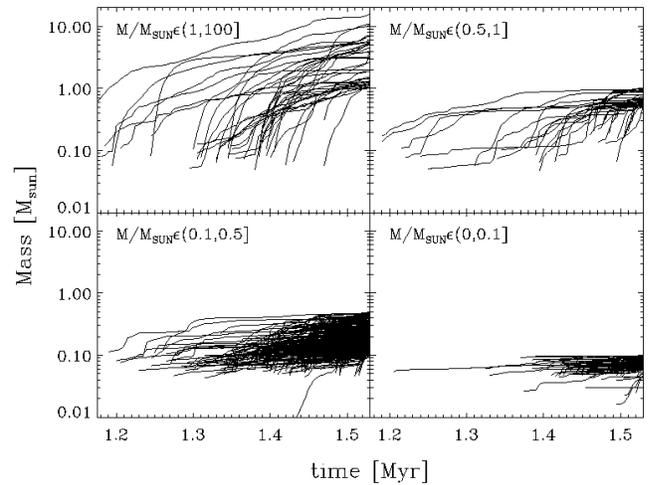}
\caption{Mass growth {\it vs.} time for sinks in run 32, separated
  into final mass bins for clarity. Upper left, sinks with eventual
  masses $> 1 \msun$; upper right, sinks with final masses $1.0 < M
  \leq 0.5 \msun$; lower left, $0.1 < M \leq 0.5 \msun$; lower right,
  $M \leq 0.1 $ sinks.}
\label{fig:masshist32}
\end{figure}

Since individual groups show the quadratic dependence when we take the
ratio $\dot{M}/\alpha^\prime$, why does the plot $\dot{M}$ {\it vs.}
$M$ for all the groups exhibits a shallower slope?  Should not they
all collapse to a single $\dot{M} \propto M^2$ relation, maybe with
substantial scatter, when plotting $\dot{M}$ vs. $M$? As shown in
Fig.~\ref{fig:sinktime}, where we plot the final mass of each sink vs.
the time at which every sink is formed, the most massive sinks
systematically form earlier than the lowest-mass sinks, and thus
accrete for longer.  In other words, {\em the integration of $\dot{M}$
  over time has an implicit dependence on mass}.  This allows the mass
function to evolve to $\Gamma = -1$.  Thus, as time proceeds, the {\it
  local} environment becomes depleted as it goes into sinks, with the
result that earlier-formed sinks accrete somewhat more slowly than
when they were created.  This environmental depletion is translated in
the shallower slopes of individual sink mass evolution {\it vs.}  time
(see Figure \ref{fig:masshist32}) than predicted by equation
(\ref{eq:masstime}) with $\alpha^\prime=$cst\footnote{Note, however,
  that by imposing a decreasing $\alpha(t)$ as a function of time, one
  can find individual mass accretion histories similar to those shown
  in Fig.\ \ref{fig:masshist32}.}.  To further support our conclusion,
we note that the simulations of \citet{hsu10}, where all sinks were
formed at the same time and continuing sink creation was not allowed,
do show $\mdot \propto M^2$ for the massive sinks.

\begin{figure}
\centering
\includegraphics[width=0.5\textwidth]{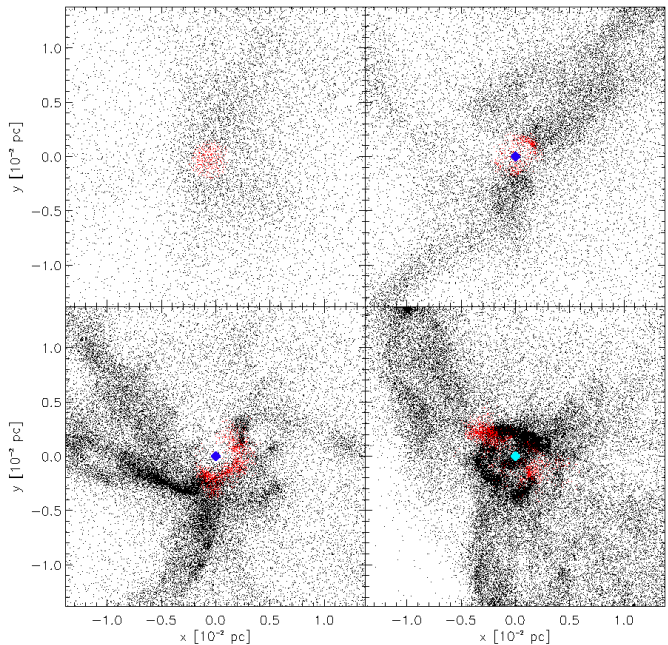}
\caption{Zoom-in of the time evolution of a region in run 22 which
  produces one of the earliest, and ultimately most massive, sinks.
  The sink (when formed) is indicated by the large symbol at (0,0).
  The red particles are those that will be accreted during the next
  time step.  The ``hole'' around the sink corresponds to the region
  $R < R(inner)$, inside of which it is assumed all gas particles
  accrete.  Gravitationally-driven infall via streams is evident (see
  text).}
\label{fig:initialsink}
\end{figure}

The self-gravitating nature of the medium, unlike that envisaged in
the simple BHL calculation, is responsible for the typically rapid
mass growth at early times, as in the ``collapse'' mode identified by
M14.  It is not surprising that this is qualitatively different than
that predicted by equation (\ref{eq:masstime}), as this early
evolution represents the initial formation of the gravitating body,
and, as such, is not addressed by the BHL formula, which assumes the
presence of an already existing mass.  One must also recognize that
the initial sink mass is only the central condensation of a larger
self-gravitating clump.  In essence, the early evolution is that of
protostellar core collapse; the initial protostellar core has a much
lower mass than its surrounding envelope, but the latter is bound by
its own self-gravity, not that of the protostellar core.  The lack of
applicability of gravitational focusing at early stages, however, does
not mean that gravitational focusing or quasi-BHL accretion cannot
occur later to build the mass well beyond initial collapse values.

One may ask how this differs from models in which turbulence produces
an initial clump mass that sets the final mass, even for massive
objects \citep[e.g.,][]{padoan02, hennchab08,hennchab09}.  We
illustrate the difference in Figure \ref{fig:initialsink}, where we
show the evolution of a region around a sink that forms early and
becomes massive.  In the upper left panel, the red particles are those
that will become part of the initial sink in the next time step (the
sink has not yet formed).  As described in the previous paragraph, the
initial sink mass is only a fraction of the larger self-gravitating
clump.  As time proceeds, the added matter arises from
gravitationally-driven streams which do not represent initial
turbulent fragments.

We note in passing that even though the ``velocity fluctuation''
models were initiated with a supersonic turbulent velocity field, the
motions rapidly decay, leaving density fluctuations as the dominant
features which then produce gravitationally-induced supersonic
motions.  Thus, it may be that initial clump masses are also
influenced by gravitational focusing.

\subsection{Comparison with other investigations}
\label{sec:comparisons}

As noted in the Introduction, in contrast to our findings, M14
concluded that ``the upper power law tail of the IMF is unrelated to
Bondi-Hoyle accretion'', and, instead, argued that the mass function
is the result of a distribution of (initial) seed masses, growth
times, and stochastic, fluctuating accretion rates with $\dot{M}
\propto M^{-2/3}$.  We suggest that the primary reason for the
differing conclusions is that their mass function is considerably
different from ours.  The mass function of M14 (their Figure 11) shows
a much broader peak than ours; inspection of their Figure 7 shows that
most of the objects in the broad peak of the mass function correspond
to ``collapse-dominated'' sinks, with the mass accreted in only one or
two episodes.  These objects arguably correspond to ``initial''
fragmentation masses, and so BHL accretion is undoubtedly irrelevant
for these objects (see previous section).  However, we also note that
the M14 mass function at masses above $1 \msun$ shows an approximate
power-law slope of $\Gamma \sim -1.35$.  M14 label these ``strong''
accretors or ``accretion-dominated'' systems, and we suggest that
BHL-type accretion is actually occuring in this mass range.

The broad peak in the M14 mass function relative to ours may reflect,
in part, the piecewise barotropic equation of state used by
\citet{bonnell08}, which can affect fragmentation (see discussion in
\cite{jappsen05}).  Our adoption of an isothermal equation of state
means that the peaks of our low-mass mass functions are essentially
set by our resolution limits rather than by any thermal physics; this
likely expands the dynamic range in mass over which BHL accretion can
operate, resulting in a clearer development of the Zinnecker power law
in our calculations.

As argued in the previous section, the apparent weaker dependence of
accretion rate on mass than $\propto M(sink)^2$ does not rule out BHL
accretion because of the depletion of environmental gas along with
competition between sinks.  In fact, when accounting for local
variations of $\alpha$, the quadratic dependence emerges.  While M14
obtain an even slower dependence of mass accretion on sink mass than
we do, this is partly the result of anchoring the fit at masses below
$1 \msun$ (their Figure 6), where accretion beyond the initial
fragmentation mass is small.

Finally, M14 also called attention to rapid, large fluctuations in
mass accretion rates as a possible problem with BHL accretion; such
variations have been seen in other simulations as well
\citep{peters10,peters11,padoan14}.  We also see strong variations in
$\mdot$ as well, but this variability does not prevent the development
of the $\Gamma = -1$ power law.  These accretion rate variations
reflect clumping in the near-sink environments (see Figure
\ref{fig:initialsink}).  It appears that these clumps are regions
which are attempting to fragment and gravitationally collapse, but
which fail to do so before being accreted.  In our runs, additional
sink formation near existing sinks is prohibited to avoid numerical
problems and short time steps as new sinks begin to orbit each other
closely.  As a test, we reran one of the simulations for a short
period of time, removing this constraint, and indeed found additional
sinks would form.  Thus, it is not clear whether the ``ingestion'' of
this material really represents accretion of gas that will be accreted
into the central object, or really represents fragmentation into an
additional sink that will simply end up orbiting the first sink.  (Of
course there is a general issue in this type of simulation, in that
resolution limits mean that we are only reporting ``system'' or
``enclosed masses within some radius''.)  In our case, the assumption
of isothermality probably enhances the tendency toward gravitational
clumping.  Thus, even though highly variable accretion may be of great
interest for understanding the time-variability of protostellar
accretion \citep[e.g.][]{audard14}, further careful consideration of
missing effects such as stellar feedback, magnetic fields, etc.\ is
needed before applying these results to real systems.

\section{Conclusions}

Our numerical experiments with simplified physics indicate that
gravitational focusing is an effective and plausible means for
producing an upper mass power law $\Gamma \sim -1$.  The situation is
more complicated than in the simplistic analysis of BHL accretion by
\citet{zinnecker82}, but even the temporal and spatial variability,
and the presence of nearby gravitating masses, does not prevent the
development of the upper mass power law.  Although the details of
accretion {\it vs.} time and sink mass are complicated, we argue that
the relative accretion rates, intergrated over time and space, exhibit
an $M^2$ dependence as in BHL accretion, which produces the power-law
mass function we find.

The present set of simulations are obviously quite idealized.  Our
purpose was to elucidate the basic physics of gravitationally-focused
accretion in as easily-visualized and interpretable a situation as
possible.  While effects of magnetic fields, non-isothermal behavior,
and stellar feedback are clearly important to any realistic and
comprehensive understanding of the stellar mass function, the results
shown here strongly suggest that BHL accretion, understood in its most
general sense as an ``attractor'' with an $M^2$ dependence on relative
accretion rates, remains relevant to an understanding of the upper
mass region of the IMF, as previously argued by
\citet{bonnell01a,bonnell01b,clark07}.

\section{Acknowledgments}

We acknowledge an extremely helpful referee report from Ian Bonnell
which helped clarify our understanding of the simulation results.
This work was supported by UNAM-PAPIIT grant number IN103012 to JBP,
and by the University of Michigan, and the Rackham graduate school.
Calculations were performed in the supercomputer at DGTIC-UNAM and at
the University of Michigan. NPG acknowledges a scholarship from
CONACYT.  We have made extensive use of the NASA-ADS database.

\label{lastpage}
\end{document}